\documentclass[preprintnumbers,superscriptaddress,showkeys,showpacs,byrevtex]{revtex4}
\usepackage{amsmath,amsfonts,amssymb,amscd,amsxtra,amsthm}
\usepackage{graphicx,bm}
\begin{document}
\preprint{PKNU-NuHaTh-2017-01}
\preprint{INHA-NTG-01/2017}
\title{Photoproduction of $\Lambda^*(1405)$ with the $N^*$ and the $t$-channel Regge contributions}
\author{Sang-Ho Kim}
\email[E-mail: ]{sangho.kim@apctp.org}
\affiliation{Asia Pacific Center for Theoretical Physics (APCTP), Pohang 37673, Republic of Korea}
\author{Seung-il Nam}
\email[E-mail: ]{sinam@pknu.ac.kr}
\affiliation{Department of Physics, Pukyong National University (PKNU), 
 Busan 608-737, Republic of Korea}
\affiliation{Asia Pacific Center for Theoretical Physics (APCTP),
 Pohang 37673, Republic of Korea}
\author{Daisuke Jido}
\email[E-mail: ]{jido@tmu.ac.jp}
\affiliation{Department of Physics, Tokyo Metropolitan University, Hachioji, Tokyo 192-0397, Japan}
\author{Hyun-Chul Kim}
\email[E-mail: ]{hchkim@inha.ac.kr}
\affiliation{Department of Physics, Inha University, Incheon 402-751,
 Republic of Korea}
\affiliation{School of Physics, Korea Institute for Advanced Study 
 (KIAS), Seoul 130-722, Republic of Korea}
\date{\today}
\begin{abstract}
We investigate the photoproduction of the $\Lambda(1405)\equiv\Lambda^*$ hyperon resonance, i.e., $\gamma p\to K^+\Lambda^*$, employing the effective Lagrangian approach with the $t$-channel Regge trajectories at tree level. We extensively explore the effects from the nucleon resonances in the vicinity of the threshold $\sqrt{s}_\mathrm{th}\approx1900$ MeV, i.e., $N^*(2000)$, $N^*(2030)$, $N^*(2055)$, $N^*(2095)$, and $N^*(2100)$, and observe that they are of great importance to reproduce the recent CLAS experimental data. Total and differential cross sections are given as numerical results and compared with the experimental data, in addition to the photon-beam asymmetry. The invariant-mass distributions for $\gamma p\to K^+\pi^0\Sigma^0$ via $\Lambda^*$ are also extracted from the two-body  process results, showing a qualitative agreement with the data. We also discuss the constituent-counting rule for the internal structure of $\Lambda^*$, resulting in that $\Lambda^*$ appears to be different from a simple three-quark $(uds)$ state. 
\end{abstract}
\maketitle

\section{Introduction}
\label{SecI}
The structure of exotic hadrons, such as the tetraquarks, pentaquarks, and meson-baryon molecular states for instance, has been one of the most interesting topics over decades in terms of the strongly interacting systems, governed by quantum chromodynamics (QCD). Recent discoveries of those exotics can shed light on the new understanding of QCD at low energies. The mesons consisting of four quarks, i.e., tetraquark state, has been reported by the Belle collaboration and BESIII collaboration~\cite{Choi:2003ue,Abe:2007jna,Choi:2007wga,Belle:2011aa,Ablikim:2013mio,Liu:2013dau,Ablikim:2013wzq}. The LHCb collaboration observed signals for the heavy pentaquark state $P_c^+$ as well~\cite{Aaij:2015tga}. The meson-baryon molecular state for $\Lambda(1405)\equiv\Lambda^*$,  rather than a simple three-quark ($uds$) one, was proposed first even before QCD was established by Dalitz and Tuan~\cite{Dalitz:1959dn,Dalitz:1960du} and recently its properties have been investigated via the unitarized chiral dynamics~\cite{Kaiser:1995eg,Oset:1997it,Nacher:1998mi,Oller:2000fj,Jido:2002zk,Jido:2003cb,Nam:2003ch,Lutz:2004sg,Magas:2005vu,Hyodo:2011ur,Mai:2014xna} and supported recently by the lattice-QCD (LQCD) simulation by investigating the strange form factor of $\Lambda^*$~\cite{Nemoto:2003ft}. The LQCD simulation also supports the meson-baryon molecular nature by investigating the strange form factor of  $\Lambda^*$~\cite{Nemoto:2003ft}. In addition to the studies of structure for the $\Lambda(1405)$, the relevant production mechanisms  were investigated extensively as well in Refs.~\cite{Nam:2008jy,Nakamura:2013boa,Nam:2015yoa,Wang:2016dtb}.

In the present work, we would like to investigate the photoproduction of $\Lambda^*$, i.e.,  $\gamma p \to K^+\Lambda^*$, employing the effective Lagrangian approach with the $K$ and  $K^*$ Regge trajectories at tree level. We focus on the contributions from the nucleon resonances near the  threshold, such as $N^*(2000,5/2^+)$ and $N^*(2100,1/2^+)$, which have been reported in the Particle Data Group (PDG)~\cite{Olive:2016xmw}. In addition to them, a few missing resonances, i.e., $N^*(2030,1/2^-)$, $N^*(2055,3/2^-)$, and $N^*(2095,3/2^-)$, predicted by the relativistic SU$(6)$ quark model~\cite{Capstick:1992uc,Capstick:1998uh}, are also taken into account.  The electromagnetic and strong couplings are determined from the presently available theoretical and experimental results. Especially, we used the chiral unitary model (ChUM) for the strong couplings for $\Lambda^*$~\cite{Khemchandani:2011mf}, because we do not have much information for $g_{KN\Lambda^*}$ and $g_{K^*N\Lambda^*}$ from experiments. The couplings for the $KN^*\Lambda^*$ vertex are taken from the quark-model calculations~\cite{Capstick:1998uh}. In order to satisfy the Ward-Takahashi (WT) identity, we make use of the gauge-invariant prescription for the form factors in the invariant amplitude. We compute various physical observables: The total $(\sigma_{\gamma p \to K^+\Lambda^*})$ and differential $(d\sigma_{\gamma p \to K^+\Lambda^*}/d\cos\theta,\,d\sigma_{\gamma p \to K^+\Lambda^*}/dt)$ cross sections, the photon-beam asymmetry $(\Sigma_{\vec{\gamma} p \to K^+\Lambda^*})$, the invariant mass plot $(d\sigma_{\gamma p \to K^+\pi^0\Sigma^0}/dM_{\pi^0\Sigma^0})$ for $\Lambda^*$, and so on. 

From the numerical results, we observe that the nucleon-resonance contributions are crucial to  reproduce  the experimental data from the CEBAF Large Acceptance Spectrometer at Jefferson laboratory (CLAS/Jlab)~\cite{Moriya:2013hwg} for the total cross section near the  threshold. Among the resonances, we find that $N^*(2000,5/2^+)$ and $N^*(2100,1/2^+)$ dominate the threshold region. As for the differential cross sections as a function of the outgoing $K^+$ angle ($\theta$) in the center-of-mass (c.m.) frame, the nucleon resonances play an important role to produce the strength of the cross sections below $\sqrt{s}\equiv W\lesssim2.2$ GeV as expected. As the production energy increases, the conventional nonresonant contributions dominate and enhance the forward peaking at $\cos\theta\approx0$, due to the strong $K$-exchange contribution.  

The $t$-dependent differential cross sections $(d\sigma_{\gamma p \to K^+\Lambda^*}/dt)$ is computed for $W=(2.0 - 4.0)$ GeV, with the help of the Regge approach, which can extend a simple low-energy Born approximation into the higher energy beyond the resonance region~\cite{Donachie2002}. As expected, the curves are obviously modified by the $N^*$ contributions near threshold. At the same time, the photon-beam asymmetry is computed as a function of $\cos\theta$ for different energies and is found to be in the shape of a distorted $\sin2\theta$ according to the competing $K$- and $K^*$-Regge contributions.

Assuming that the decay width of $\Lambda^*$ is sufficiently narrow $(\Gamma_{\Lambda^*}\approx50\,\mathrm{MeV}\ll \Lambda_\mathrm{hadron}\approx1\,\mathrm{GeV})$ and the interference between the $\Lambda^*$ and other nonstrange mesons decaying into $K\bar{K}$ are negligible in the Dalitz process of $\gamma p\to K^+\pi^0\Sigma^0$, the differential cross section of $d\sigma_{\gamma p\to K^+\pi^0\Sigma^0}/dM_{\pi^0\Sigma^0}$ can be obtained from a simple formula with $\Gamma_{\Lambda^*\to\pi\Sigma}$ and
$\sigma_{\gamma p\to K^+\Lambda^*}$, which is computed previously~\cite{Nam:2015yoa}. By doing this, the distribution as a function of invariant mass $M_{\pi^0\Sigma^0}$ is drawn and compared with the data, showing a good agreement and supporting its successful application.  

Finally, we investigate the constituent-counting rule (CCR). The CCR is a method to analyze the internal structure of the hadrons involved in the $2\to2$ reaction process by dimensional considerations of the reaction amplitude in terms of the quark and gluon propagators at the large angle as well as the high energy. Applying this to the present reaction process, we observe that the numerical result, i.e., $s^7d\sigma/dt$ as a function of $W$, differs clearly from the three-quark state for $\Lambda^*$, although it does not lead to the concrete conclusion that the results support the five-quark state for $\Lambda^*$.

The present work is organized as follows: In Sec.~\ref{SecII}, theoretical  framework is briefly explained. Numerical results and relevant discussions are given in Sec.~\ref{SecIII}. The final section is devoted for summary.
\section{Theoretical Framework}
\label{SecII}
\begin{figure}[t]
\centering
\includegraphics[width=15cm]{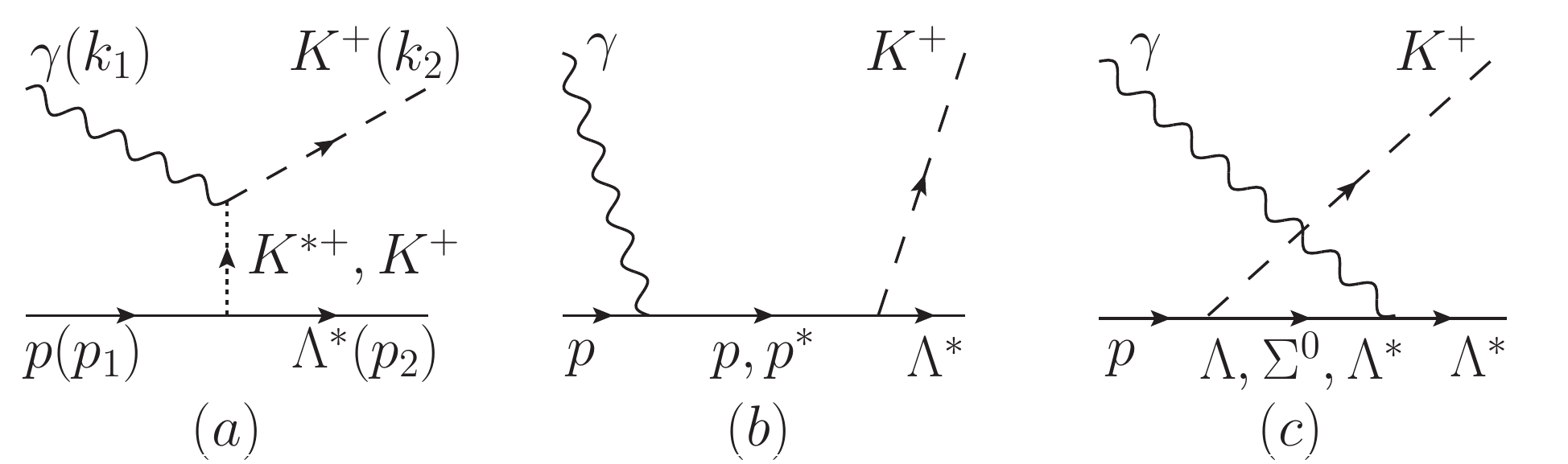}
\caption{Relevant tree-level Feynman diagrams for the $\gamma p \to K^+ \Lambda(1405)$ reaction process. $p^*$, $\Lambda$, $\Sigma^0$, and $\Lambda^*$ stand for the proton resonances, $\Lambda(1116)$, $\Sigma^0(1193)$, and $\Lambda(1405)$, respectively.}
\label{FIG1}
\end{figure}

In this section, we provide a brief explanation for the present theoretical framework. Basically, we employ the tree-level Born approximation with effective Lagrangians for the interaction vertices and the Regge trajectories for the pseudoscalar (PS) and vector (V) meson exchanges in the $t$ channel. In terms of the PS meson-baryon coupling scheme, the relevant Feynman diagrams for the $\gamma p \to K^+ \Lambda^*(1405)$ reaction process are drawn in Fig.~\ref{FIG1}, in which $k_1$ and $p_1$ stand for the four momenta for the incident photon and target proton, whereas $k_2$ and $p_2$ for the outgoing $K^+$ and recoiled $\Lambda^*(1405)$, respectively. (a) As for the $t$-channel, we consider the $K$ and $K^*$ exchanges with their {\it Regge} propagators. By doing this, we can explore higher energy regions, which cannot be probed by the simple Born approximation~\cite{Donachie2002}. (b) The nucleon and its resonance states are taken into account for the baryon-pole diagrams in the $s$ channel, whereas (c) the hyperons, $\Lambda(1116)$, $\Sigma^0(1193)$, and $\Lambda(1405)$, are included for the $u$ channel as shown in Fig.~\ref{FIG1}. Although there are other possible hyperon contributions in the $u$ channel, such as $\Lambda(1520)$ and $\Lambda(1670)$ for instance, the magnetic transitions to $\Lambda(1405)$ have not been reported experimentally as well as theoretically. In addition, the $u$-channel resonances with a higher mass do not produce significant structure in the cross section in the energy region we are interested in. Hence, we will not take those higher-mass hyperons into account in the present calculation.

The effective Lagrangians for the EM interaction vertices read
\begin{eqnarray}
\mathcal L_{\gamma K K} &=& 
-ie_K [  K^\dagger (\partial_\mu K) - (\partial_\mu K^\dagger) K ] A^\mu ,   
 \cr  
\mathcal L_{\gamma K K^*} &=&
g_{\gamma K K^*} \epsilon^{\mu\nu\alpha\beta}
\partial_\mu A_\nu
[ (\partial_\alpha K_\beta^{*-}) K^+ + K^- (\partial_\alpha K_\beta^{*+} )] , 
 \cr
\mathcal L_{\gamma NN} &=&
- \bar N \left[ e_N \gamma_\mu - \frac{e\kappa_N}{2M_N}
\sigma_{\mu\nu}\partial^\nu \right] A^\mu N ,                           
 \cr   
\mathcal L_{\gamma \Lambda^* \Lambda^*} &=&
\frac{e\mu_{\Lambda^*}}{2M_N}
\bar \Lambda^* \sigma_{\mu\nu} \partial^\nu A^\mu \Lambda^* ,
\cr
\mathcal L_{\gamma Y \Lambda^*} &=&
\frac{e\mu_{\Lambda^* \to Y \gamma }}{2M_N}
\bar Y \gamma_5 \sigma_{\mu\nu} \partial^\nu A^\mu \Lambda^* 
+ \mathrm{H.c.},
\label{eq:BornLag1}
\end{eqnarray}
where $A_\mu$, $K$, $K^*$, $N$, and $\Lambda^*$ indicate the fields for the photon, pseudoscalar kaon, vector kaon, nucleon, and $\Lambda(1405)$, respectively. $Y$ corresponds to the field for the ground-state $\Lambda$ or $\Sigma^0$. $M_h$ and $e_h$ stand for the mass and electric charge of the hadron $h$, while $e$ denotes the unit electric charge. As for the values for the coupling constants, the charged $g_{\gamma K K^*}^c$ is calculated from the experimental data for the decay width $\Gamma(K^* \to K \gamma)$, resulting in 0.254\,$\mathrm{GeV}^{-1}$~\cite{Olive:2016xmw}. The anomalous magnetic moment of the proton is given by $\kappa_N =1.79$. The transition magnetic moments between two hyperons are also necessary and given by $\mu_{h\to h'}$. The SU(3) quark model gives  $\mu_{\Lambda^*}=0.44$~\cite{Williams:1991tw} which is lied within the values predicted from the ChUM: $\mu_{\Lambda^*}=0.2 - 0.5$~\cite{Jido:2002yz}. Meanwhile, we obtain $\mu_{\Lambda^*\to(\Lambda,\Sigma^0)\gamma}=(-0.43,0.61)$ from an isobar model~\cite{Burkhardt:1991ms} to match the $K^- p$  atom data~\cite{Whitehouse:1989yi}. Thus, the output of the radiative decay widths is given by $\Gamma_{\Lambda^* \to \gamma \Lambda} = (27 \pm 8)$ keV and $\Gamma_{\Lambda^* \to \gamma \Sigma^0} = (23 \pm 7)$  keV~\cite{Burkhardt:1991ms} from the formula
\begin{eqnarray}
\Gamma_{\Lambda^* \to \gamma Y} 
=\frac{(e\mu_{\Lambda^* \to \gamma Y})^2{\bf k}^3}{4\pi M^2_N},
\label{DW:GYLs}
\end{eqnarray}
derived from the $\gamma Y \Lambda^*$ Lagrangian in Eq.~(\ref{eq:BornLag1}). Here {\bf k} is the magnitude of the three-momentum of the hyperon $Y$ in the rest frame of $\Lambda^*$.

The effective Lagrangians for the strong vertices are written by
\begin{eqnarray}
\mathcal L_{K N Y} &=&
- i g_{K N Y} \bar N \gamma_5 Y K + \mathrm{H.c.},         
 \cr
\mathcal L_{K N\Lambda^*} &=& 
-i g_{K N \Lambda^*} \bar N \Lambda^* K + \mathrm{H.c.},
 \cr
\mathcal L_{K^* N \Lambda^*} &=&-
g_{K^* N \Lambda^*} \bar N\gamma_5 \gamma_\mu \Lambda^* K^{*\mu}
+ \mathrm{H.c.}.
\label{eq:BornLag2}
\end{eqnarray}
Note that the strong coupling $g_{K N(\Lambda,\Sigma^0)}$ is given by $(-13.4, 4.09)$ from the Nijmegen soft-core potential (NSC97a)~\cite{Stoks:1999bz}. Because there is no sufficient experimental information on the strong coupling constants for the excited $\Lambda$ hyperons, we resort to theoretical results, using the chiral-unitary model (ChUM)~\cite{Khemchandani:2011mf}. Averaging those theoretical values for various cases, we determine the strengths for the couplings for the numerical calculations as $|g_{KN\Lambda^*}|\simeq 1.95$ and $|g_{K^*N\Lambda^*}|\simeq 1.3$. All the values for the relevant couplings for the numerical calculations are summarized in Table~\ref{TAB1}.

\begin{table}[b]
\begin{tabular}{|c|c|c|c|c|c|c|c|c|}
\hline
$g_{\gamma K K^*}^c$
&$\kappa_N$
&$\mu_{\Lambda^*}$
&$\mu_{\Lambda^*\to\gamma\Lambda}$
&$\mu_{\Lambda^*\to\gamma\Sigma^0}$
&$g_{K N \Lambda}$
&$g_{ K N \Sigma^0}$
&$|g_{K N \Lambda^*}|$
&$|g_{K^* N \Lambda^*}|$\\
\hline
$-0.254/\mathrm{GeV}$
&$1.79$~\cite{Olive:2016xmw}
&$0.44$~\cite{Williams:1991tw}
&$-0.43$~\cite{Burkhardt:1991ms,Whitehouse:1989yi} 
&$0.61$~\cite{Burkhardt:1991ms,Whitehouse:1989yi}
&$-13.4$~\cite{Stoks:1999bz}
&$4.09$~\cite{Stoks:1999bz}
&$1.95$~\cite{Khemchandani:2011mf}
&$1.3$~\cite{Khemchandani:2011mf}\\
\hline
\end{tabular}
\caption{Relevant EM and strong coupling constants for the numerical calculations.}
\label{TAB1}
\end{table}

The invariant-scattering amplitude for the photoproduction can be written in general by  
\begin{eqnarray}
\mathcal{M} = I_h \bar u_{\Lambda^*} \mathcal{M}^\mu_h \epsilon_\mu u_N,
\label{eq:AmpNotation}
\end{eqnarray}
where $u_N$ and $u_{\Lambda^*}$ designate the Dirac spinors for the target nucleon and recoiled $\Lambda ^*$, respectively, and $\epsilon_\mu$ denotes the polarization vector of the incident photon. In the present calculation, the isospin factors are given by $I_K = I_N = I_{K^*}= I_\Lambda = I_{\Sigma^0} = I_{\Lambda^*} =  1$.  The effective Lagrangians of Eqs.~(\ref{eq:BornLag1}) and (\ref{eq:BornLag2}) being employed, the relevant hadronic amplitude ($\mathcal{M}^\mu_h$) besides the nucleon-resonance ($N^*$) contributions can be obtained straightforwardly as follows:
\begin{eqnarray}
\mathcal M_K^\mu&=& -2ie g_{K N \Lambda^*} \frac{1}{t-M_K^2} k_2^\mu ,     
 \cr
\mathcal M_N^\mu&=& -ie g_{K N \Lambda^*}
\frac{\rlap{/}{q_s}+M_N}{s-M_N^2}
\left [ \gamma^\mu  + \frac{i\kappa_p}{2M_N} 
\sigma^{\mu\nu} k_{1\nu} \right ] ,                                     
 \cr
\mathcal M_{K^*}^\mu&=& g_{\gamma K K^*} g_{K^* N \Lambda^*} 
\frac{1}{t-M_{K^*}^2} \epsilon^{\mu\nu\alpha\beta} 
\gamma_5 \gamma_\nu k_{1\alpha} k_{2\beta} ,                              
 \cr
\mathcal M_{\Lambda,\Sigma^0}^\mu&=&
\frac{e\mu_{\Lambda^*\to{\gamma(\Lambda,\Sigma^0)}}}{2M_N}
\frac{g_{K N (\Lambda,\Sigma^0)}}{u-M_{(\Lambda,\Sigma^0)}^2}
\sigma^{\mu\nu} k_{1\nu}
(\rlap{/}{q_u}-M_{(\Lambda,\Sigma^0)}) ,
 \cr
\mathcal M_{\Lambda^*}^\mu&=&
\frac{e\mu_{\Lambda^*}}{2M_N}
\frac{g_{K N \Lambda^*}}{u-M_{\Lambda^*}^2}
\sigma^{\mu\nu} k_{1\nu}
(\rlap{/}{q_u}+M_{\Lambda^*}) ,
\label{eq:BornEachAmp}
\end{eqnarray}
where $q_{s,t,u}$ stand for the off-shell four momenta, defined by $q_s = k_1+p_1$, $q_t=k_1-k_2$, and $q_u = p_2-k_1$, and we also have $q^2_{s,t,u}=(s,t,u)$, which denote the Mandelstam variables. 

Considering the spatial extension of the hadrons, one needs to take into account the empirical form factors in the numerical calculations. We introduce a form factor as follows:
\begin{equation}
F = F (x)=
\left[ \frac{\Lambda^4}{\Lambda^4+\left(x-M^2\right)^2} \right]^2.
\label{eq:FF}                                                       
\end{equation}
Here, $x$ and $\Lambda$ indicate the Mandelstam variables and hadronic cutoff-mass parameter. Because the naive usage of the form factors can violate the Ward-Takahashi (WT) identity, various effective prescriptions to preserve the identity were suggested in Refs.~\cite{Ohta:1989ji,Haberzettl:1997jg,Haberzettl:1998eq,Davidson:2001rk,Haberzettl:2015exa}. The prescription from Ref.~\cite{Davidson:2001rk} being followed, the bare invariant amplitude is reconstructed with the form factors, satisfying the WT identity, as follows:
\begin{eqnarray}
\epsilon\cdot\mathcal M_{\mathrm{Born}}&=&
\epsilon\cdot\left[\left(\mathcal M_K + \mathcal M_N\right) \,F_c
+\mathcal M_{K^*}\, F_{K^*}                 
+\mathcal M_\Lambda\, F_\Lambda
+\mathcal M_{\Sigma^0}\,F_{\Sigma^0}
+\mathcal M_{\Lambda^*}\,F_{\Lambda^*} \right],
\label{eq:BornAmp}
\end{eqnarray}
where we define a common form factor as
\begin{eqnarray}
F_c = F_{t,K} + F_{s,N} - F_{t,K}F_{s,N},
\label{eq:CFF}
\end{eqnarray}
which fulfills the on-mass-shell condition, i.e., the form factor becomes unity at $q^2_x=m^2_h$, and the crossing symmetry. It is easy to verify that $k_\gamma\cdot\mathcal M_{\mathrm{Born}} = k_1\cdot\mathcal M_{\mathrm{Born}}=0$, i.e., satisfying the WT identity, as already shown in the previous works~\cite{Nam:2008jy,Nam:2015yoa}.

Now, we are in a position to consider the $N^*$ contributions in the $s$ channel. As shown in the previous work~\cite{Nam:2015yoa}, the nucleon resonance was found to be crucial to reproduce the experimental data. In the present work, we consider more resonances in the vicinity of the reaction threshold. Among the nucleon resonances listed in the Particle Data Group (PDG)~\cite{Olive:2016xmw}, we take into account $N^*(2000,5/2^+)$ and $N^*(2100,1/2^+)$, which are near the reaction threshold and couple strongly to $\gamma N$~\cite{Olive:2016xmw} as well as $K\Lambda^*$~\cite{Capstick:1998uh} channels. Other resonances, for instance, $N^*(1895,1/2^-)$, $N^*(1900,3/2^+)$, $N^*(1990,7/2^+)$, and $N^*(2060,5/2^-)$, are excluded from our consideration, because their couplings to the $K\Lambda^*$ channel are very small or even  exhibit zero values ~\cite{Capstick:1998uh}. Meanwhile, the SU (6) relativistic-quark model provides us with missing resonances, such as  $N^*(2030,1/2^-)$, $N^*(2055,3/2^-)$, and $N^*(2095,3/2^-)$~\cite{Capstick:1998uh,Capstick:1992uc}. Hence, five $N^*$ states in total are taken into account in the present work. For this purpose, we first define the effective Lagrangians for the EM transitions for them with respect to their spin and parity $(j^P)$ as follows:
\begin{eqnarray}
\mathcal{L}^{1/2^\pm}_{\gamma  N N^*} &=& 
\frac{eh_1}{2M_N} \bar N \Gamma^{\mp}
\sigma_{\mu\nu} \partial^\nu A^\mu N^* + \mathrm{H.c.} ,               
\cr
\mathcal{L}^{3/2^\pm}_{\gamma N N^*}&=& 
-ie \left[ \frac{h_1}{2M_N} \bar N \Gamma_\nu^{\pm}
 - \frac{ih_2}{(2M_N)^2} \partial_\nu \bar N
 \Gamma^{\pm} \right] F^{\mu\nu} N^*_\mu + \mathrm{H.c.},   
 \cr 
\mathcal{L}^{5/2^\pm}_{\gamma N N^*} &=&
e\left[ \frac{h_{1}}{(2M_N)^2} \bar N \Gamma_\nu^{\mp}
-\frac{ih_{2}}{(2M_N)^3} \partial_\nu \bar N
\Gamma^{\mp} \right] \partial^\rho F^{\mu\nu} 
N^*_{\mu\rho} + \mathrm{H.c.} ,   
\label{eq:ResLag1}
\end{eqnarray}
where $h^{J}_i$ denotes the EM transition coupling obtained from the Breit-Wigner helicity amplitudes $A^{J}_i$~\cite{Olive:2016xmw}. The explicit relations between them are given in Refs.~\cite{Oh:2007jd,Oh:2011}. Similarly, the effective Lagrangians for the strong interactions read:
\begin{eqnarray}
\mathcal{L}^{1/2^\pm}_{K \Lambda^* N^*}&=& 
 - i g_{K \Lambda^* N^*} \bar K \bar \Lambda^* 
 \Gamma^{\mp} N^* + \mathrm{H.c.},                                   
 \cr
\mathcal{L}^{3/2^\pm}_{K \Lambda^* N^*}&=& 
 \frac{g_{K \Lambda^* N^*}}{M_K} \partial^\mu \bar K \bar \Lambda^* 
 \Gamma^{\pm} N^*_\mu + \mathrm{H.c.},                            
 \cr  
\mathcal{L}^{5/2^\pm}_{K \Lambda^* N^*} &=& 
 \frac{ig_{K \Lambda^* N^*}}{M_K^2} \partial^\mu \partial^\nu \bar K 
 \bar \Lambda^* \Gamma^{\mp} N^*_{\mu\nu} + \mathrm{H.c.},
\label{eq:ResLag2}
\end{eqnarray}  
where $N^*$, $N^*_\mu$, and $N^*_{\mu\nu}$ denote the spin-1/2, -3/2, and -5/2 nucleon-resonance fields, respectively. To construct the nucleon resonances, whose spins are greater than $1/2$, one needs a special description, such as the Rarita-Schwinger (RS) formalism~\cite{Rarita:1941mf,Nath:1971wp}. In this formalism, there appear some theoretical uncertainties. In Refs.~\cite{David:1995pi,Mizutani:1997sd,Vrancx:2011qv,Deser:2000dz}, the authors explored them by addressing the gauge invariant RS fields, off-shell effects, causality, etc. Although these subjects are interesting to study, it must be beyond our scope for the present research. Therefore, we make use of the simplest prescription for the RS fields as done in our previous work~\cite{Kim:2011rm,Kim:2012pz}. Note that we utilized the following notations depending on the parities of
$N^*$s ($P=\pm$):
\begin{eqnarray}
\Gamma^{\pm} = \left(
\begin{array}{c} 
\gamma_5 \\ I_{4\times4}
\end{array} \right) ,
\,\,\,\,
\Gamma_\mu^{\pm} = \left(
\begin{array}{c}
\gamma_\mu \gamma_5 \\ \gamma_\mu 
\end{array} \right) .
\label{eq:GammaPM}
\end{eqnarray}
Using Eqs.~(\ref{eq:ResLag1}) and (\ref{eq:ResLag2}), it is straightforward to compute the invariant amplitudes for the nucleon-resonance contributions with $\mathcal M = I_{N^*} \bar u_{\Lambda^*} \mathcal{M}_{N^*} u_N$ as in Eq.~(\ref{eq:AmpNotation}):
\begin{eqnarray}
\mathcal M^{1/2^\pm}_{N^*}&=& 
\mp g_{K \Lambda^* N^*} \frac{eh_1}{2M_N} 
 \frac{\Gamma^{\mp}(\rlap{/}{q_s} + M_{N^*})\Gamma^{\mp}}
{s-M_{N^*}^2+iM_{N^*}\Gamma_{N^*}} 
 \sigma^{\mu\nu} k_{1\nu} \epsilon_\mu ,                      
\cr
\mathcal M^{3/2^\pm}_{N^*}&=& 
i \frac{g_{K \Lambda^* N^*}}{M_K}
 \frac{\Gamma^{\pm}   k_2^\mu }
 {s-M_{N^*}^2+iM_{N^*}\Gamma_{N^*}} 
\Delta_{\mu\nu} (q_s)                                      
\left[ \frac{eh_1}{2M_N} \Gamma_\lambda^{\pm} \mp 
       \frac{eh_2}{(2M_N)^2} \Gamma^{\pm} p_{1\lambda} \right]
(k_1^\nu \epsilon^\lambda - k_1^\lambda \epsilon^\nu) ,                       
 \cr
\mathcal M^{5/2^\pm}_{N^*}&=& 
i \frac{g_{K \Lambda^* N^*}}{M_K^2}
\frac{\Gamma^{\mp}   k_2^{\mu} k_2^{\nu}}
{s-M_{N^*}^2+iM_{N^*}\Gamma_{N^*}} 
\Delta_{\mu \nu}^{\rho \sigma} (q_s)                     
\left[ \frac{eh_1}{(2M_N)^2} \Gamma_\lambda^{\mp} \pm 
       \frac{eh_2}{(2M_N)^3} \Gamma^{\mp} p_{1\lambda} \right]
k_{1\sigma} (k_{1\rho} \epsilon^\lambda - k_1^\lambda \epsilon_{\rho}) ,  
\label{eq:ResEachAmp}
\end{eqnarray}
where $\Gamma_{N^*}$ stands for the full decay width for $N^*$. The spin-summation factor for spin-$3/2$ and spin-$5/2$ spinors are assigned by $\Delta_{\mu\nu}$ and $\Delta_{\mu\nu}^{\rho\sigma}$ and their explicit forms given by~\cite{Berends:1979rv,Oh:2007jd,Oh:2011}
\begin{eqnarray}
\Delta_{\mu\nu}(q)&=&(\rlap{/}{q} + M_{N^*})
\left[ - g_{\mu\nu} +\frac{1}{3} \gamma_\mu \gamma_\nu  +
  \frac{1}{3M_{N^*}} (\gamma_\mu q_\nu - \gamma_\nu q_\mu)
 +\frac{2}{3M^2_{N^*}} q_\mu q_\nu \right],
 \cr
\Delta_{\mu\nu}^{\rho\sigma}(q)&=&(\rlap{/}{q} + M_{N^*})
 \left[\frac{1}{2}( \bar g_{\mu}^{\rho} \bar g_{\nu}^{\sigma} 
                   +\bar g_{\mu}^{\sigma} \bar g_{\nu}^{\rho})
-\frac{1}{5} \bar g_{{\mu}{\nu}} \bar g^{{\rho}{\sigma}}
-\frac{1}{10}(
 \bar \gamma_{\mu}\bar \gamma^{\rho} \bar g_{\nu}^{\sigma}  
+\bar \gamma_{\mu}\bar \gamma^{\sigma} \bar g_{\nu}^{\rho} 
+\bar \gamma_{\nu}\bar \gamma^{\rho} \bar g_{\mu}^{\sigma}
+\bar \gamma_{\nu}\bar \gamma^{\sigma} \bar g_{\mu}^{\rho}) 
\right].
\label{eq:RSSP}
\end{eqnarray}
Here, we have used the following notations for convenience:
\begin{equation}
\bar{g}_{\mu\nu} = g_{\mu\nu} - \frac{q_\mu q_\nu}{M_{N^*}^2}, 
\,\,\,\,
\bar{\gamma}_\mu = \gamma_\mu - \frac{q_\mu}{M_{N^*}^2}\rlap{/}{q}.
\end{equation}

The relevant input parameters from the PDG and missing nucleon resonances are summarized in Table~\ref{TAB2}. The EM transition ($h^{J}_i$) and strong coupling $g_{K\Lambda^*N^*}$) constants are derived from the experimental~\cite{Olive:2016xmw} and theoretical information~\cite{Capstick:1998uh,Capstick:1992uc}. Note that we adopt the central ones among the values $A_i^J$ and $G(\ell)$. The details for obtaining the strong coupling constants, $g_{K\Lambda^*N^*}$, are explained in the Appendix. Because the phase factors between the invariant  amplitudes for different $N^*$s cannot be determined simply by symmetries, such as the gauge and flavor symmetries for instance, it is natural for them to be considered as free parameters to reproduce the data. In general, those amplitudes are represented by
\begin{eqnarray}
\mathcal{M}_{\mathrm{Res}} = 
\sum_{N^*} e^{i\psi_{N^*}} {\mathcal M_{N^*}} F_{N^*},
\label{eq:ResAmp}
\end{eqnarray}
where $\psi_{N^*}$ is a certain phase angle and $F_{N^*}$ indicates the form factor, whose form is the same as $F(s)$ of Eq.~(\ref{eq:FF}). We note that, in Refs.~\cite{Corthals:2005ce,Corthals:2006nz,DeCruz:2012bv}, in order to suppress the nucleon-resonance contributions in the high-energy regions, the Gaussian form factors were employed. In describing the data, we, however, verified that our choice for the $s$-channel form factor in Eq.~(\ref{eq:FF}) works sufficiently to reproduce the data as shown in Sec.~\ref{SecIII}. We reach a similar conclusion when the following Gaussian form factor is instead used with the same cutoff mass
$\Lambda_{N^*}$ = 0.9 GeV:
\begin{eqnarray}
F_{\mathrm{Gauss}}(s) = \mathrm{exp}
\left\{ - \frac{(s-M_{N^*}^2)^2}{\Lambda_{N^*}^4} \right\}.
\label{eq:GaussFF}
\end{eqnarray}

\begin{table}[b]
\begin{tabular}{|c|cccc|cc|}
\hline
&$A_{1/2}$
&$A_{3/2}$
&$h_1$
&$h_2$
&$G(\ell)$~\cite{Capstick:1998uh}
&$g_{K \Lambda^* N^*}$\\
\hline
$N^*(2000,5/2^+)$~\cite{Olive:2016xmw}
&$31\pm 10$&$-43 \pm 8$&$-4.22$&$3.98$
&$-0.6_{-1.6}^{+0.6}$&$-0.912$ \\
$N^*(2100,1/2^+)$~\cite{Olive:2016xmw}
&$10\pm 4$&$...$&$-0.045$&$...$  
&$+5.2 \pm 0.8$&$0.785$ \\ 
$N^*(2030,1/2^-)$~\cite{Capstick:1992uc}
&$20$&$...$&$0.094$&$...$&$+1.2_{-1.1}^{+0.9}$&$1.78$ \\
$N^*(2055,3/2^-)$~\cite{Capstick:1992uc}
&$16$&$0$&$-0.335$&$0.419$&$+1.2_{-0.9}^{+0.5}$&$-0.467$ \\
$N^*(2095,3/2^-)$~\cite{Capstick:1992uc}
&$-9$&$-14$&$0.018$&$-0.134$&$+0.7_{-0.4}^{+0.2}$&$-0.228$ \\
\hline
\end{tabular}
\caption{The input parameters for the nucleon-resonance contributions. The helicity amplitudes $A_{1/2,\,3/2}$ [$10^{-3}/\sqrt{\mathrm{GeV}}$] are obtained from Refs.~\cite{Olive:2016xmw,Capstick:1992uc} and the decay amplitudes $G(\ell)$ [$\sqrt{\mathrm{MeV}}$] are extracted from Ref.~\cite{Capstick:1998uh}.}
\label{TAB2}
\end{table}

Although we would like to reproduce the data in the relatively low-energy region near the threshold, as done in the CLAS experiment, it is interesting to explore the higher-energy region theoretically for future experiments in the upgraded CLAS/Jlab and other experimental facilities. For this purpose, we employ the $t$-channel Regge trajectories for $K$ and $K^*$ mesons and follow closely Ref.~\cite{Guidal:1997hy}. In this Regge approach, the Feynman propagators in Eq.~(\ref{eq:BornEachAmp}) are replaced simply by the Regge ones as
\begin{eqnarray}
\label{eq:REGGEPRO}
\frac{1}{t-M_K^2} &\to& P_K^{\mathrm{Regge}}=                      
\nonumber
\left( \frac{s}{s_0} \right)^{\alpha_K}
\frac{\pi\alpha_K'}{\sin(\pi\alpha_K)}
\left\{ \begin{array}{c} 1 \\ e^{-i\pi\alpha_K} \end{array} \right\}
\frac{1}{\Gamma(1+\alpha_K)}, \\ 
\frac{1}{t-M_{K^*}^2} &\to& P_{K^*}^{\mathrm{Regge}}=
\left( \frac{s}{s_0} \right)^{\alpha_{K^*}-1}
\frac{\pi\alpha'_{K^*}}{\sin(\pi\alpha_{K^*})}
\left\{ \begin{array}{c} 1 \\ e^{-i\pi\alpha_{K^*}} \end{array} \right\}
\frac{1}{\Gamma(\alpha_{K^*})},       
\end{eqnarray}
where the Regge phases can be the constant $(1)$ or the rotating $( e^{-i\pi\alpha_{K,K^*}})$ one. The Regge trajectories read~\cite{Guidal:1997hy}
\begin{eqnarray}
\alpha_K=\alpha_K(t)&=& \frac{0.7}{\mathrm{GeV}^2} (t - M_K^2), \,\,\,\,  
\alpha_{K^*}=\alpha_{K^*}(t)= \frac{0.83}{\mathrm{GeV}^2}t + 0.25 ,
\label{eq:ReggeTraj}
\end{eqnarray}
and we define the slope parameter as $\alpha'_{K,K^*}\equiv\partial\alpha_{K,K^*}(t)/\partial t$. Conventionally, the energy scale parameter is chosen to be $s_0 = 1\, 
\mathrm{GeV^2}$. With this in mind, the bare invariant amplitude of $t$ and $s$ channels in Eq.~(\ref{eq:BornAmp}) is modified as
\begin{eqnarray}
\epsilon\cdot\mathcal M_{t,s}^{\mathrm{Regge}}&=&
\epsilon\cdot
\left[\left(\mathcal M_K + \mathcal M_N\right) (t - M_K^2) 
P_K^{\mathrm{Regge}} \,
+\mathcal M_{K^*} (t - M_{K^*}^2) P_{K^*}^{\mathrm{Regge}}
\right],
\label{eq:BornAmp2}
\end{eqnarray}
for the present reaction process.

It is worth mentioning that the three-body (Dalitz) reaction process, $ab\to cde$, can be explored approximately in terms of the two-body one, if one assumes the following: (1) The decay widths for the decaying resonances are sufficiently narrow and (2) the interference between the different resonances in the Dalitz plot is negligible. If these conditions are fulfilled, one can write the differential cross section (invariant-mass plot) for the Dalitz process $\gamma p\to K^+\pi^0\Sigma^0$ as follows~\cite{Nam:2015yoa}:
\begin{eqnarray}
\label{eq:APP2}
\frac{d\sigma_{\gamma p\to K^+\pi^0\Sigma^0}}{dM_{\pi^0\Sigma^0}}
&\approx&\frac{2M_{\Lambda^*}M_{\pi^0\Sigma^0}}{\pi}
\frac{\sigma_{\gamma p\to K^+\Lambda^*}\,\Gamma_{\Lambda^*\to\pi^0\Sigma^0}}
{(M^2_{\pi^0\Sigma^0}-M^2_{\Lambda^*})^2+M^2_{\Lambda^*}\Gamma^2_{\Lambda^*}},
\end{eqnarray}
where $\sigma_{\gamma p\to K^+\Lambda^*}$ is the two-body total cross section. Here are some justifications for the usage of Eq.~(\ref{eq:APP2}): First, the decay width of $\Lambda^*\to \pi\Sigma$ is about $50$ MeV, which is much smaller than the typical energy range $\sim 500$ MeV in the present analysis. Therefore, condition (1) can be assumed to be reasonable. Second, in the Dalitz process $\gamma p\to K^+\pi^0\Sigma^0$, the $\Lambda^*$ production interferes with the $K^{+*}$ one on the Dalitz plot. Interestingly, in Ref.~\cite{Ryu:2016jmv}, it was found experimentally that the interference between the different resonant productions is almost negligible, although they focused on the different reactions process, i.e., $\gamma p\to K^+K^-p$. Hence, considering these observations, conditions (1) and (2) can be justified rather safely here, and we can use Eq.~(\ref{eq:APP2}) for computing the Dalitz process with the two-body cross section, which is computed by Eqs.~(\ref{eq:BornEachAmp}) and (\ref{eq:ResEachAmp}). In doing this, the absolute value for the $K^+$ three momentum is obtained as a function of the invariant mass $M_{\pi^0\Sigma^0}$ as follows:
\begin{equation}
\label{eq:MOMO}
|\vec{k}_{K^+}|=
\left[\left(
\frac{s+M^2_{K^+}-M^2_{\pi^0\Sigma^0}}
{2W}\right)^2-M^2_{K^+}\right]^\frac{1}{2}.
\end{equation}
Note that $|\vec{k}_{K^+}|=|\vec{k}_{\Lambda^*}|$ by construction in the c.m. frame.

As mentioned previously, the internal structure of $\Lambda^*$ has been one of the most interesting subjects in the nonperturbative QCD. There have been several approaches to pin down the genuine structure of the hyperon resonance. Among various theoretical studies, we discuss the constituent-counting rule (CCR)~\cite{Kawamura:2013iia,Chang:2015ioc}. Basically, in the CCR, the high-energy and large-angle scattering amplitudes for two-body processes, i.e., $ab \to cd$, are analyzed by dimensional considerations of the quark and gluon propagators, resulting in
\begin{equation}
\label{eq:CCR}
\frac{d\sigma_{ab\to cd}}{dt}\propto\frac{1}{s^{n-2}},
\end{equation}
where $n$ is the total number of the constituents of the particles involved in the scattering process. For instance, if $\Lambda^*$ is composed of three quarks, the value of $n$ becomes $1_\gamma+3_N+2_K+3_{\Lambda^*}=9$, whereas $n=11$ for the five-quark system. Note that there are many uncertain theoretical ingredients, such as the distribution functions for the involved particles and so on, to apply the CCR to real problems of $\Lambda^*$~\cite{Chang:2015ioc}. It is, however, still valuable to test the relation in Eq.~(\ref{eq:CCR}) with the present results, in which the Regge approach can extend the low-energy Born approximation to certain high energies beyond the 
resonance region.

\section{Numerical results and Discussions}
\label{SecIII}
In this section, we present the numerical results with corresponding discussions. Note that, in the present work, the free parameters are the sign of the coupling constants $g_{KN\Lambda^*}$ and $g_{K^*N\Lambda^*}$, the cutoff mass $\Lambda$ in Eq.~(\ref{eq:FF}), the phase angle $\psi_{N^*}$ in Eq.~(\ref{eq:ResAmp}), and the Regge phases in Eq.~(\ref{eq:REGGEPRO}). The cutoff masses are given in common for all the baryons as $\Lambda_{\Lambda,\Sigma,\Lambda^*,N^*}=0.9$ GeV. Although the full-decay widths for the nucleon resonance, which appear in the denominator of the amplitudes in Eq.~(\ref{eq:ResEachAmp}), are not free parameters as given in the PDG list~\cite{Olive:2016xmw}, we fix it to be $\Gamma_{N^*}=300$ MeV for all the resonances for brevity. We verified that about $\pm10\%$ deviations in the widths do not make crucial differences in the qualitative consequences of the present work. The numerical results are given by using the constant and rotating Regge phases with the fitted parameters as listed in Table~\ref{TAB3} to reproduce the CLAS/Jlab data.
\begin{table}[b]
\begin{tabular}{|c|c|c|c|}
\hline
Regge phase
&\hspace{0.5cm}$\psi_{N^*}$\hspace{0.5cm}
&\hspace{0.5cm}$g_{KN\Lambda^*}$\hspace{0.5cm}
&\hspace{0.5cm}$g_{K^*N\Lambda^*}$\hspace{0.5cm}\\
\hline
$1$ (constant)&$e^{i\pi/2}$&$1.95$&$-1.3$ \\
$e^{-i\pi\alpha_{K,K^*}}$ (rotating) &$e^{i\pi}$&$1.95$&$1.3$ \\
\hline
\end{tabular}
\caption{Fitted parameter setups the constant and rotating Regge phases. See Eqs.~(\ref{eq:ResAmp}) and (\ref{eq:REGGEPRO}) for details.}
\label{TAB3}
\end{table}

\begin{figure}[h]
\includegraphics[width=16cm]{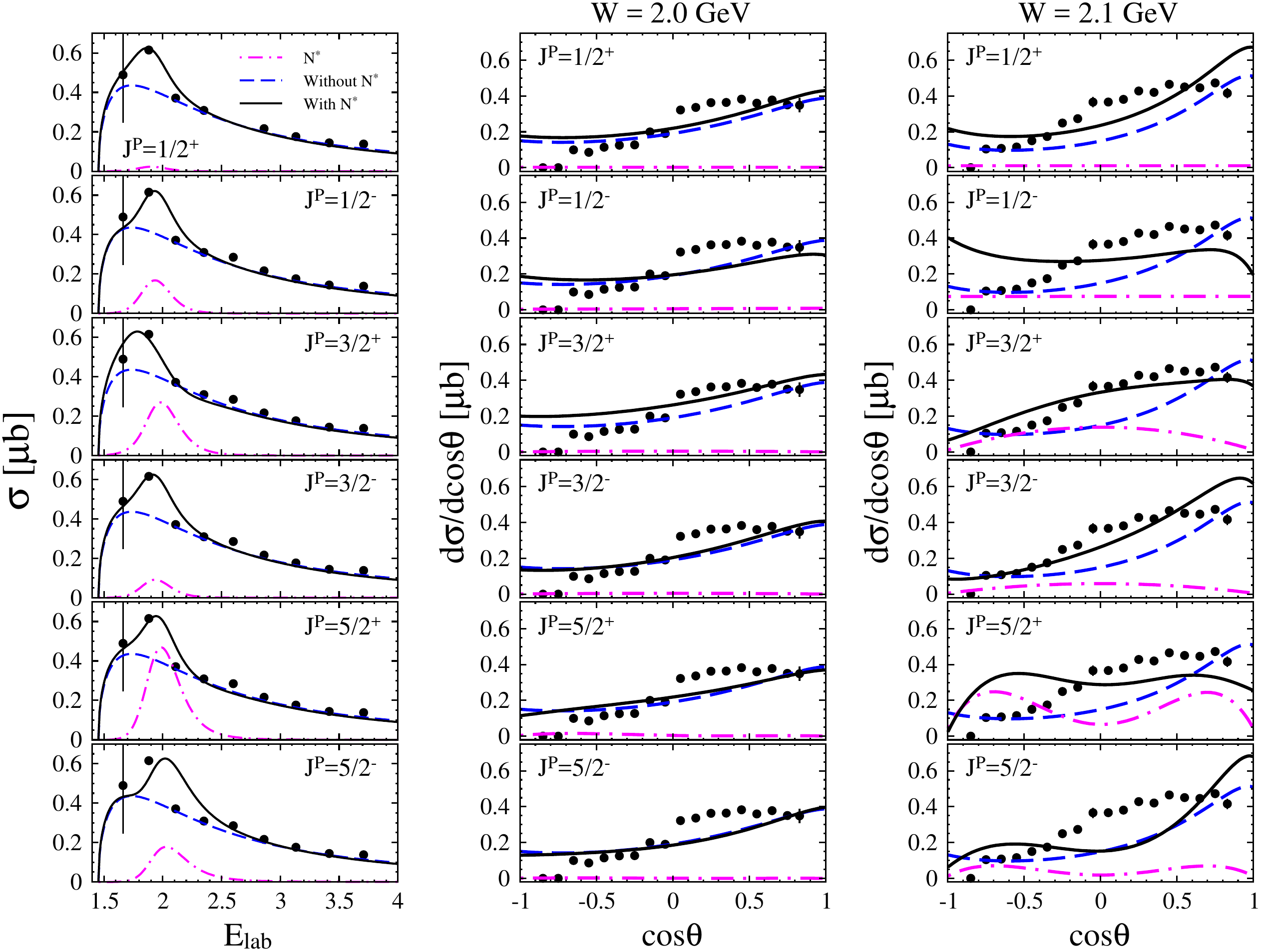}
\caption{(Color online) 
Computed total and differential cross sections for six different spins and parities for nucleon resonances with $(M_{N^*},\Gamma_{N^*})=(2.1,0.25)$ GeV.}       
\label{FIG2}
\end{figure}

Before performing the detailed calculations for reproducing the CLAS/Jlab experimental data, we would like to examine the effects of the nucleon-resonance contributions for the $\Lambda^*$ photoproduction in a model-independent manner. Here we choose the constant Regge phase for simplicity. We consider a single resonance near the reaction threshold with different spin and parity, i.e. $N^*(2100,n/2^\pm)$ for a positive integer $n\le5$. For simplicity, we choose $\Gamma_{N^*}=250$ MeV and $A_{1/2}=(3\times10^{-2})/\sqrt{\mathrm{GeV}}$ for all the resonances. The reasons for those choices are as follows: (1) In the PDG list, the reported values for $\Gamma_{N^*}$s reside in $(100 - 400)$ MeV with sizable uncertainties. Hence, the sort of a middle value is chosen. (2) Because we focus on the vicinity of the threshold, the $h_2$ contribution $\propto \partial N/M_{N^*}$ turns out to be small. Therefore, we only consider the $h_1$ contribution and the values of $A_{1/2}$ are in the order of $10^{-2}/\sqrt{\mathrm{GeV}}$ as shown in the PDG list. Thus, similarly, the middle value is employed. In Fig.~\ref{FIG2}, six resonances are taken into account in total, depending on the spins and parities, and $g_{KN^*\Lambda^*}$ is fitted with the total cross section from the CLAS/Jlab data with the fixed $h_1$ values via the above $A_{1/2}$. As a result, the branching ratios of $\mathrm{BR}_{n/2^\pm}$ are given as follows:
\begin{eqnarray}
\label{eq:GN}
&&\mathrm{BR}_{1/2^+}=1.7 \times 10^{-2},\,\,\,\,
\mathrm{BR}_{1/2^-}=0.11,\,\,\,\,\,\,\,\,\,\,\,\,\,\,\,\,\,\,\,\,
\mathrm{BR}_{3/2^+}=9.5 \times 10^{-3},              
\cr
&&\mathrm{BR}_{3/2^-}=4.0 \times 10^{-3},\,\,\,\,
\mathrm{BR}_{5/2^+}=2.1 \times 10^{-2},\,\,\,\,
\mathrm{BR}_{5/2^-}=5.9 \times 10^{-3},
\end{eqnarray}
As the spin of $N^*$ increases, the shapes of the angular distributions exhibit more fluctuations on their behaviors. We find that the nucleon resonances with $J^P$= $1/2^+$, $3/2^+$, and $3/2^-$ more or less describe the CLAS/Jlab data within reasonable values of branching ratios, where $\mathrm{BR} = \Gamma_{N^* \to K \Lambda^*}/ \Gamma_{N^*}$ and $\Gamma_{N^* \to K \Lambda^*}$ is calculated from $g_{K \Lambda^* N^*}$ for each of the cases. But when more than a single resonance is considered simultaneously, we reach a different conclusion as we will see soon.

\begin{figure}[h]
\includegraphics[width=8.0cm]{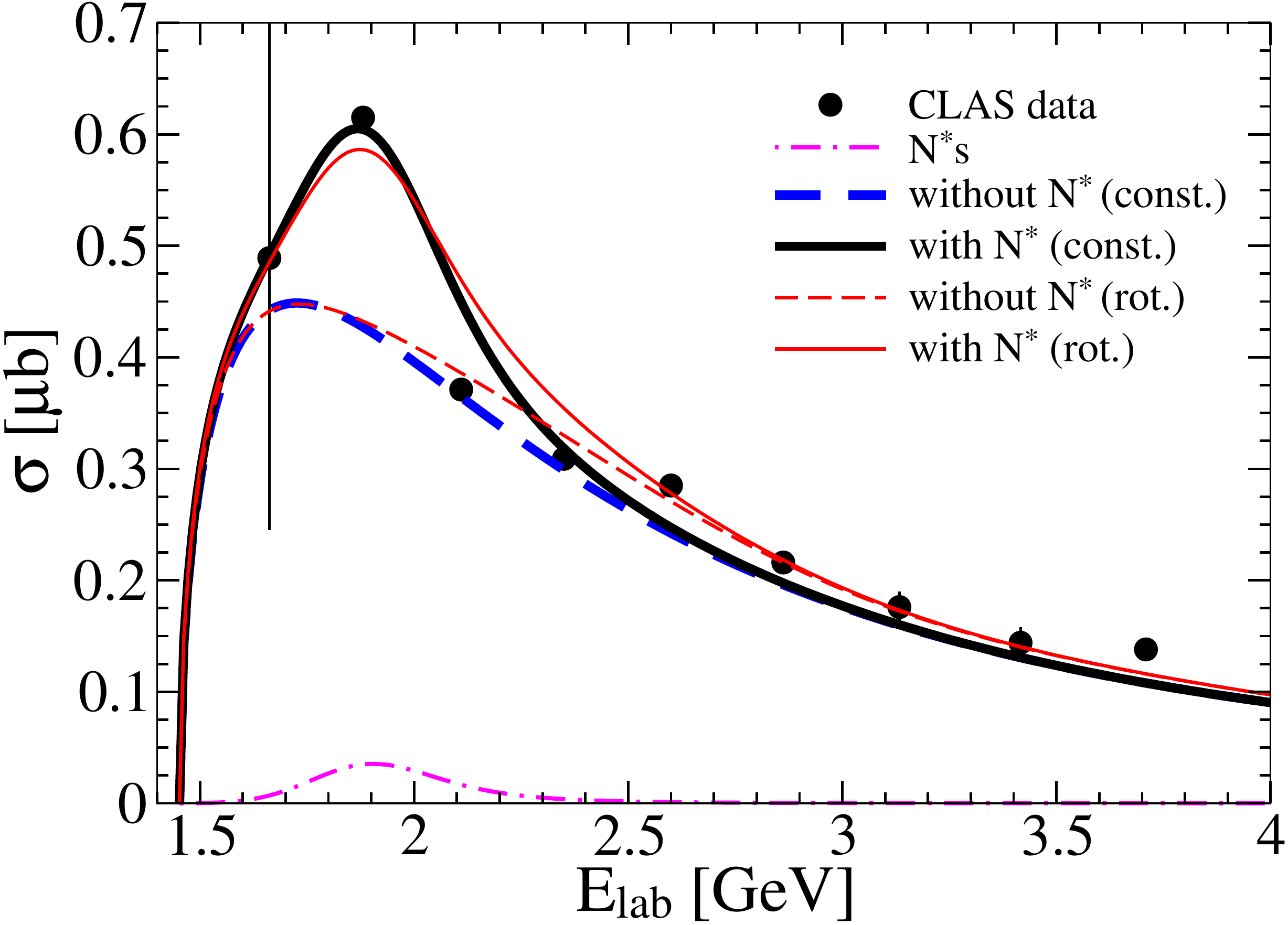}
\includegraphics[width=8.0cm]{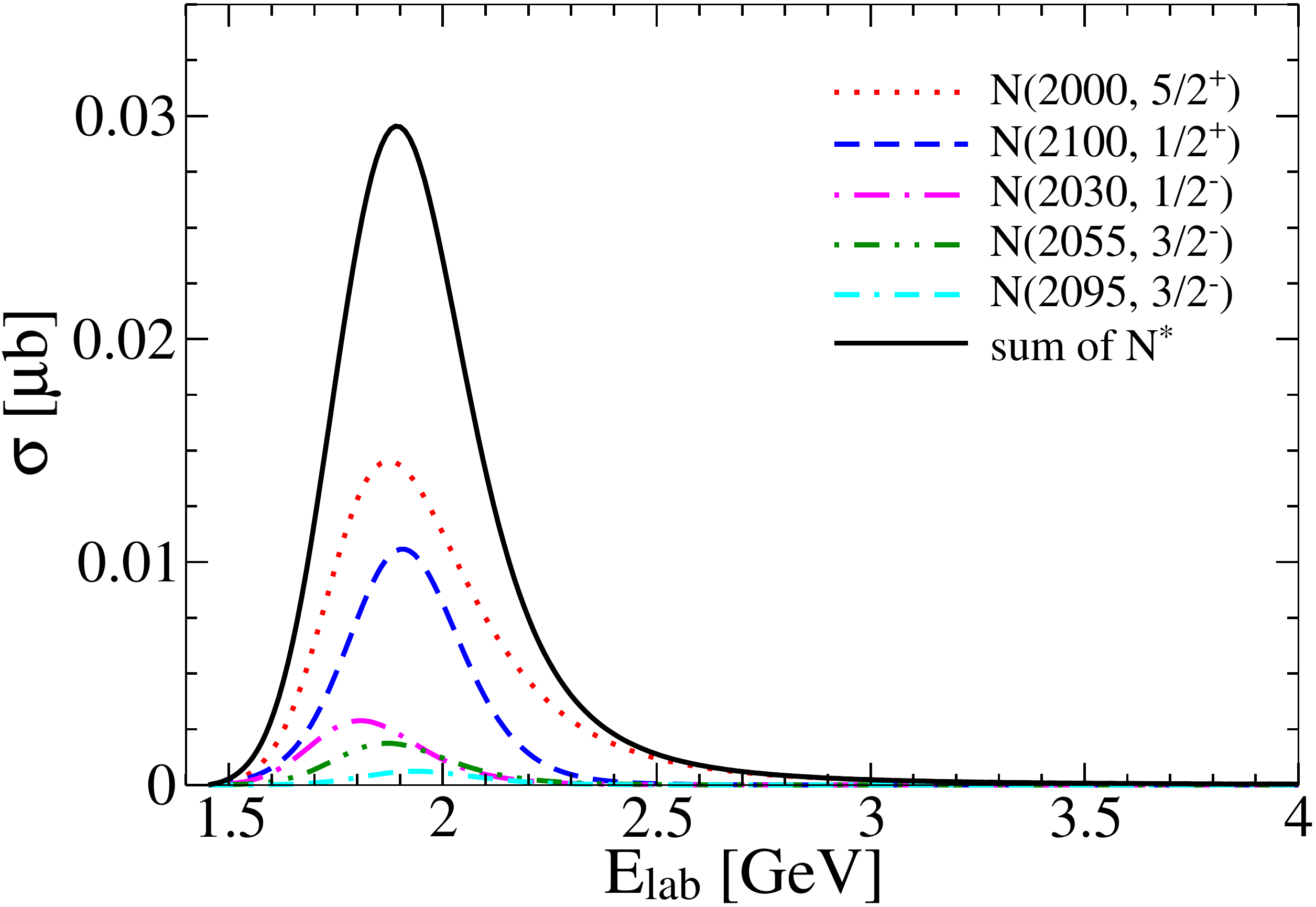}
\caption{(Color online) Left: $\sigma_{\gamma p \to K^+ \Lambda^*}$ as a function of $E_{\mathrm{lab}}$. The solid, dashed, and dot-dashed lines indicate the total, without $N^*$, and $N^*$ contributions, respectively.
The thick and thin lines correspond to the constant and rotating Regge 
phases for the $K$ and $K^*$ trajectories, respectively.
The data are taken from Ref.~\cite{Moriya:2013hwg}. Right: Separate contributions from various nucleon resonances.}
\label{FIG3}
\end{figure}

Now, let us show numerical results when employing our models. In the left panel of Fig.~\ref{FIG3}, we demonstrate the results for the total cross section for $\gamma p\to K^+\Lambda^*$ as a function of the photon laboratory energy $E_\mathrm{lab}$ by using the constant (thick lines) and rotating (thin lines) Regge phases as given in Table~\ref{TAB3}. The experimental data are taken from the CLAS/Jlab collaboration~\cite{Moriya:2013hwg}. The full result including the $N^*$ (dot-dashed) and non-$N^*$ (dashed) contributions is given in the solid line. It is worth mentioning that the peaklike bump around $E_\mathrm{lab}=2$ GeV is well reproduced by the nucleon resonances, whereas the non-$N^*$ contribution curve decreases rapidly, due to the Regge propagators, as the energy increases. We also verified that the non-$N^*$ contributions cannot reproduce the bump by changing form-factor types or their cutoff parameters. Moreover, by comparing the strengths of the $N^*$ and total contributions, one can conclude that the constructive interference between the $N^*$ and others plays an important role. As a consequence, within the most conventional tree-level approach like the present effective model, the nucleon resonances are crucial to explain the obvious cross-section enhancement near the threshold region for $\Lambda(1405)$ photoproduction. It is also interesting to see the separate effects from the different nucleon resonances. As shown there, after fitting the parameters to reproduce the data, the two Regge phases do not make obviously qualitative differences in the cross sections. Hence, in what follows, we will only use the constant Regge phase for the numerical calculations.

In the right panel of Fig.~\ref{FIG3}, we draw the total cross sections only from each resonance. It is clearly shown that $N^*(2000)$ and $N^*(2100)$ give dominant effects on the cross section, due to their relatively larger strong and EM couplings than other resonances. In contrast, the missing resonances play only minor roles in producing the strength. The strong enhancement near the threshold region was already reported from the Laser Electron Photon Experiment at Super Proton ring-8 GeV (LEPS/SPring-8)~\cite{Niiyama:2008rt}. The curve shape beyond the resonance region is reproduced mainly by the $t$-channel $K$-exchange contribution, in addition to the small but finite $K^*$-exchange one. 

\begin{figure}[h]
\includegraphics[width=15cm]{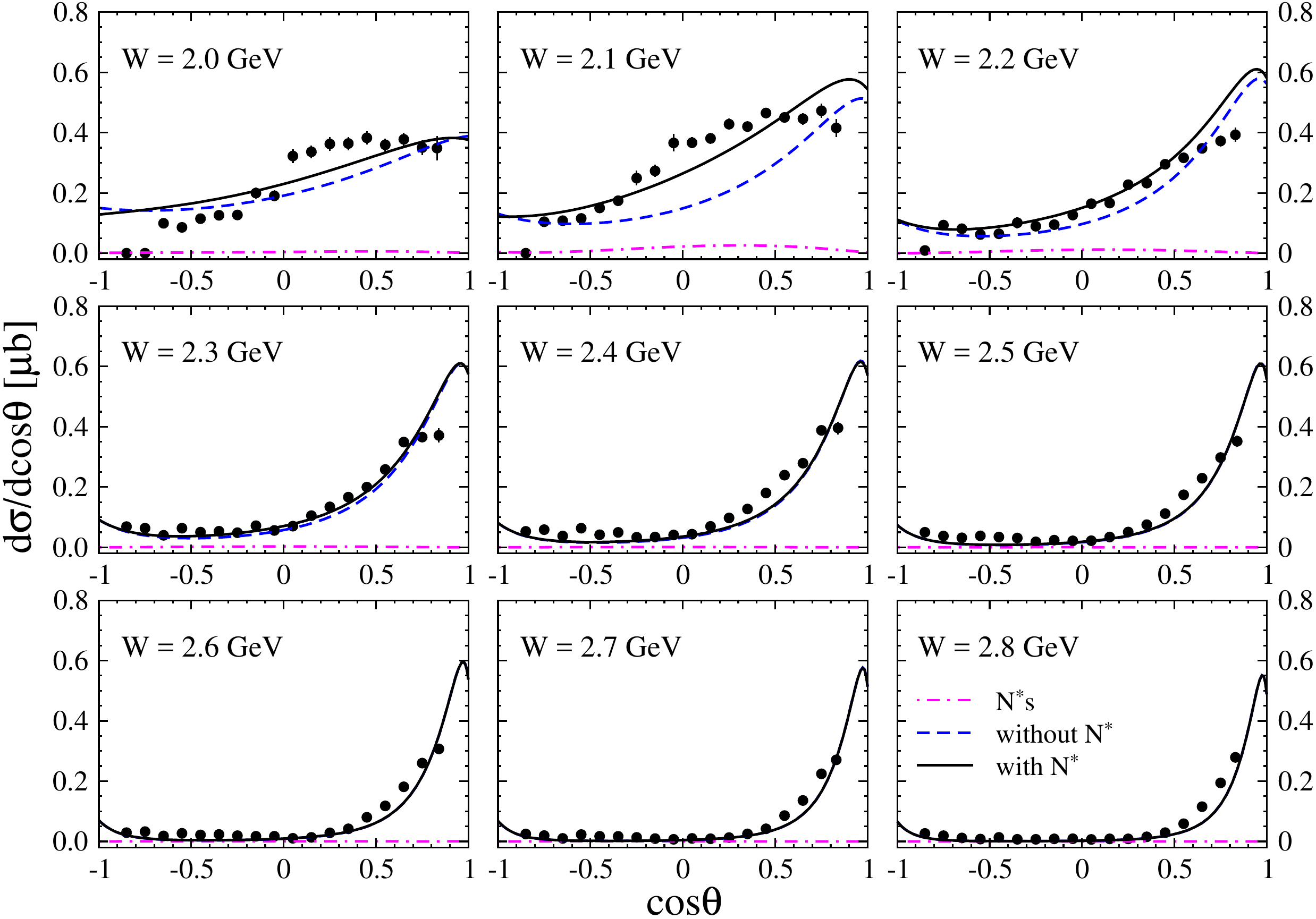} 
\caption{(Color online) $d\sigma_{\gamma p \to K^+ \Lambda^*}/d\cos\theta$ for different cm energies $W=(2.0\sim2.8)$ GeV. The data are taken from Ref.~\cite{Moriya:2013hwg}. The legends are the same with those for the left panel of
Fig.~\ref{FIG3}.}
\label{FIG4}
\end{figure}

The numerical results for the angular dependence $d\sigma/d\cos\theta$ are drawn in Fig.~\ref{FIG4} for different c.m. energies $ W=(2.0 - 2.8)$ GeV. In overall, the full results (solid) show qualitatively good agreement with the data. We also observe that the forward-scattering enhancement becomes more obvious as the energy increases, due to the strong $t$-channel $K$ exchange. At $W=2.1$ GeV, it is found that the data are reproduced by the $N^*$ and non-$N^*$ contributions constructively. This observation agrees well with that shown in the total cross section. As the energy gets higher, the $N^*$ contribution becomes diminished and almost flat as expected.  At the same time, the small but sizable backward-scattering enhancement starts appearing gradually beyond $W=2.3$ GeV. We verified that it originates mainly from the larger couplings in case of $\Lambda$ and $\Sigma^0$ exchanges, i.e., $g_{K N(\Lambda,\Sigma^0)}$ and $\mu_{\Lambda^*\to(\Lambda,\Sigma^0)}$, in the $u$ channel than other hyperons, as given in Table~\ref{TAB1}.

Note that, quantitatively, in the low-energy region $W \le 2.2$ GeV, we find sizable disagreement with the data for very backward angles $-1 \le\cos\theta\le -0.8$. Similar problems also take place in Refs.~\cite{David:1995pi,Mizutani:1997sd} for $K\Lambda$ photoproduction where the SU(3) symmetry limits on the Born coupling constants were considered. Although we anticipate that the unknown contributions beyond the present model setup can resolve it, we will not make further detailed  investigation on this issue in the present work and would like to leave it for the future. We verified that, even without our Regge approach, i.e., by using  Eq.~(\ref{eq:BornAmp}) rather than Eq.~(\ref{eq:BornAmp2}), we can obtain similar results for total and  differential cross sections in the range of $E_\mathrm{lab} \leq 4.0$ GeV. This is expected because our Regge model interpolates between low and high energy regions smoothly. 

\begin{figure}[h]
\includegraphics[width=8.0cm]{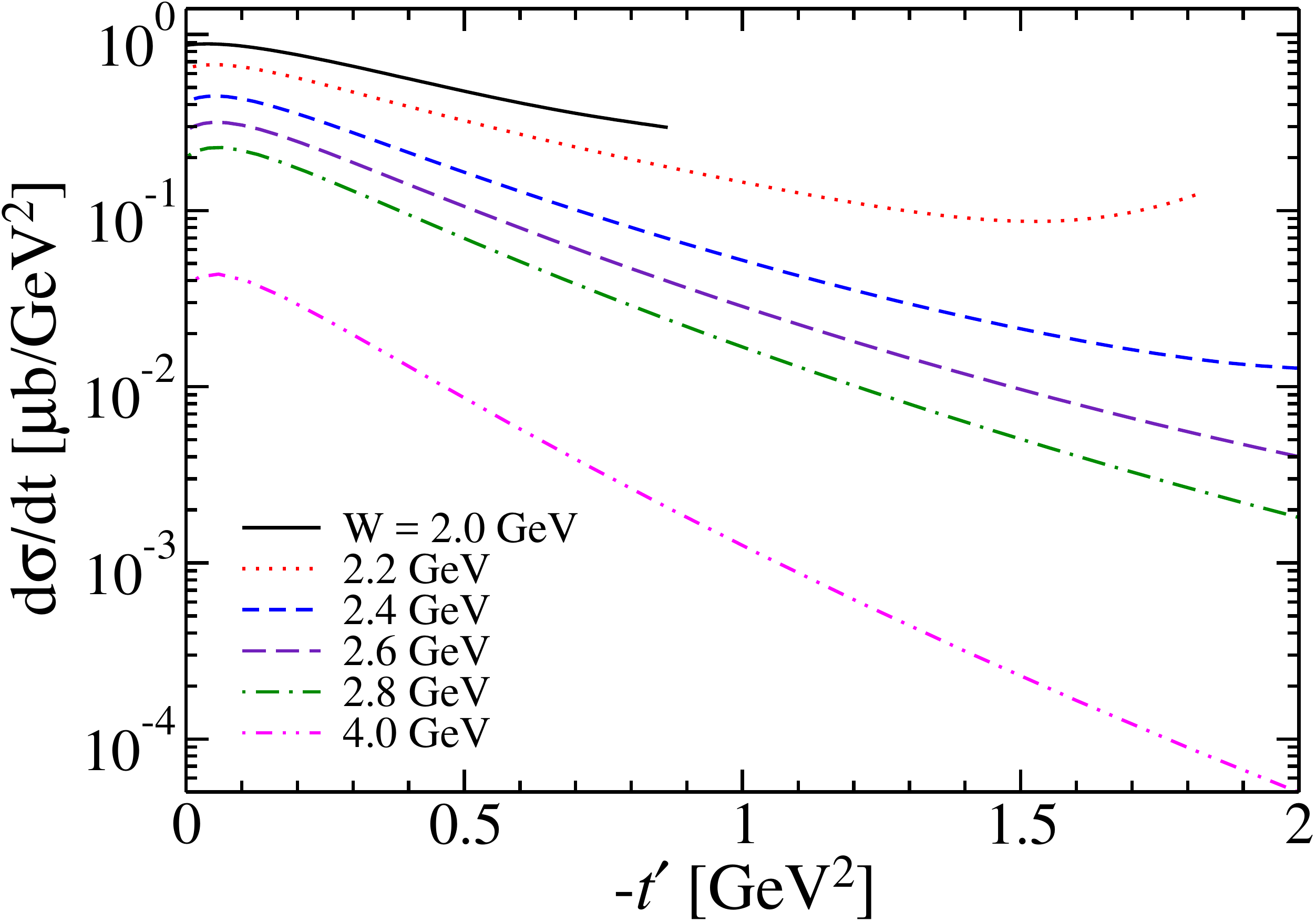}
\includegraphics[width=8.0cm]{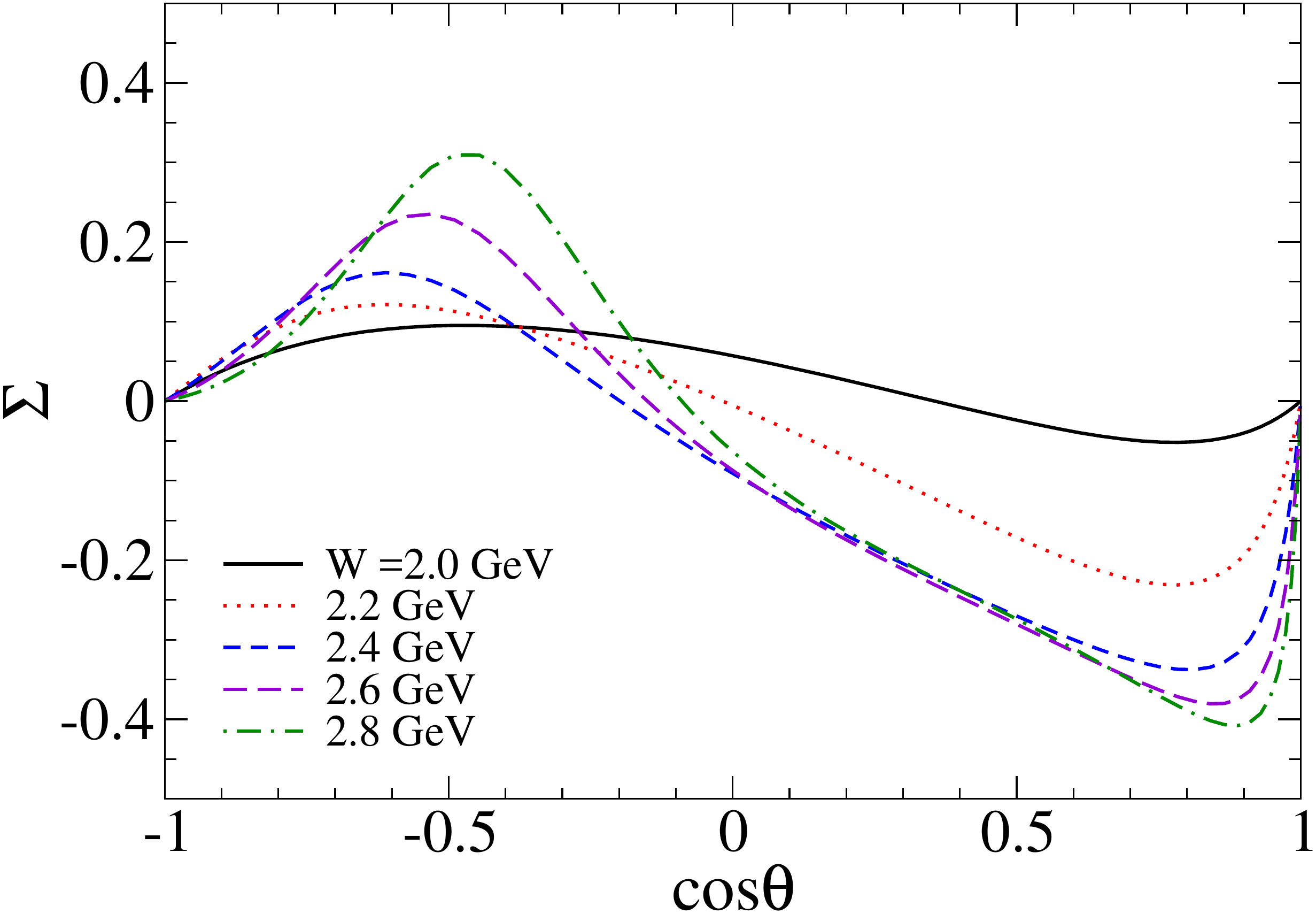}
\caption{(Color online) Left: $d\sigma_{\gamma p \to K^+ \Lambda^*}/dt$ as a function of $-t'\equiv-(t-t_\mathrm{max})$ for different cm energies $W=(2.0\sim4.0)$ GeV. Right: $\Sigma_{\vec{\gamma} p \to K^+ \Lambda^*}$ as a function of $\cos\theta$ for different cm energies $W=(2.0\sim2.8)$ 
GeV.}
\label{FIG5}
\end{figure}

In the left panel of Fig.~\ref{FIG5}, we draw the numerical results for the differential cross sections $d\sigma/dt$ as a function of $-t'\equiv-(t-t_\mathrm{max})$ for different c.m. energies $W=(2.0 - 4.0)$ GeV. The curves are computed with the $N^*$ contributions and show a typical behavior of the momentum transfer. The slope is mostly determined by the dominant $K$ Regge trajectory.
The photon-beam asymmetry, i.e., the analyzing power is one of the important physical quantities, observed in hadron photoproductions. We define it as follows:
\begin{equation}
\label{eq:BA}
\Sigma_{\vec{\gamma} p \to K^+ \Lambda^*}=\frac{\frac{d\sigma}{d\Omega}_\perp-\frac{d\sigma}{d\Omega}_\parallel}
{\frac{d\sigma}{d\Omega}_\perp+\frac{d\sigma}{d\Omega}_\parallel},
\end{equation}
where the subscripts $\parallel$ and $\perp$ stand for the cases when the photon polarization vector is parallel and perpendicular to the reaction plane, respectively. In the case of unnatural parity exchange ($K$ exchange), the perpendicular term in Eq.~(\ref{eq:BA}) is equal to zero, so $\Sigma_{\vec{\gamma} p \to K^+ \Lambda^*}= -1$, whereas natural parity exchange ($K^*$ exchange) leads $\Sigma_{\vec{\gamma} p \to K^+ \Lambda^*}$ to positive values, since the corresponding perpendicular term dominates the parallel one. In the right panel of Fig.~\ref{FIG5}, we present the numerical results for $\Sigma_{\vec{\gamma} p \to K^+ \Lambda^*}$ in the same manner with the left panel of Fig.~\ref{FIG5} for different c.m. energies $W=(2.0 - 2.8)$ GeV. Note that the beam asymmetry becomes reversed around $\cos\theta\approx0$, due to the competing $K$ and $K^*$ contributions. The effects of the $N^*$s turn out to be negligible. It is worth mentioning that the polarized quantity $\Sigma_{\vec{\gamma} p \to K^+ \Lambda^*}$ is very sensitive to the choice of the Regge phases for the $t$-channel strange-meson exchanges, because it is almost dominated by the Regge contributions. To confirm the effects of the Regge phases in the polarized physical quantities, one may need more reliable data to compare with. 

\begin{figure}[h]
\includegraphics[width=14.2cm]{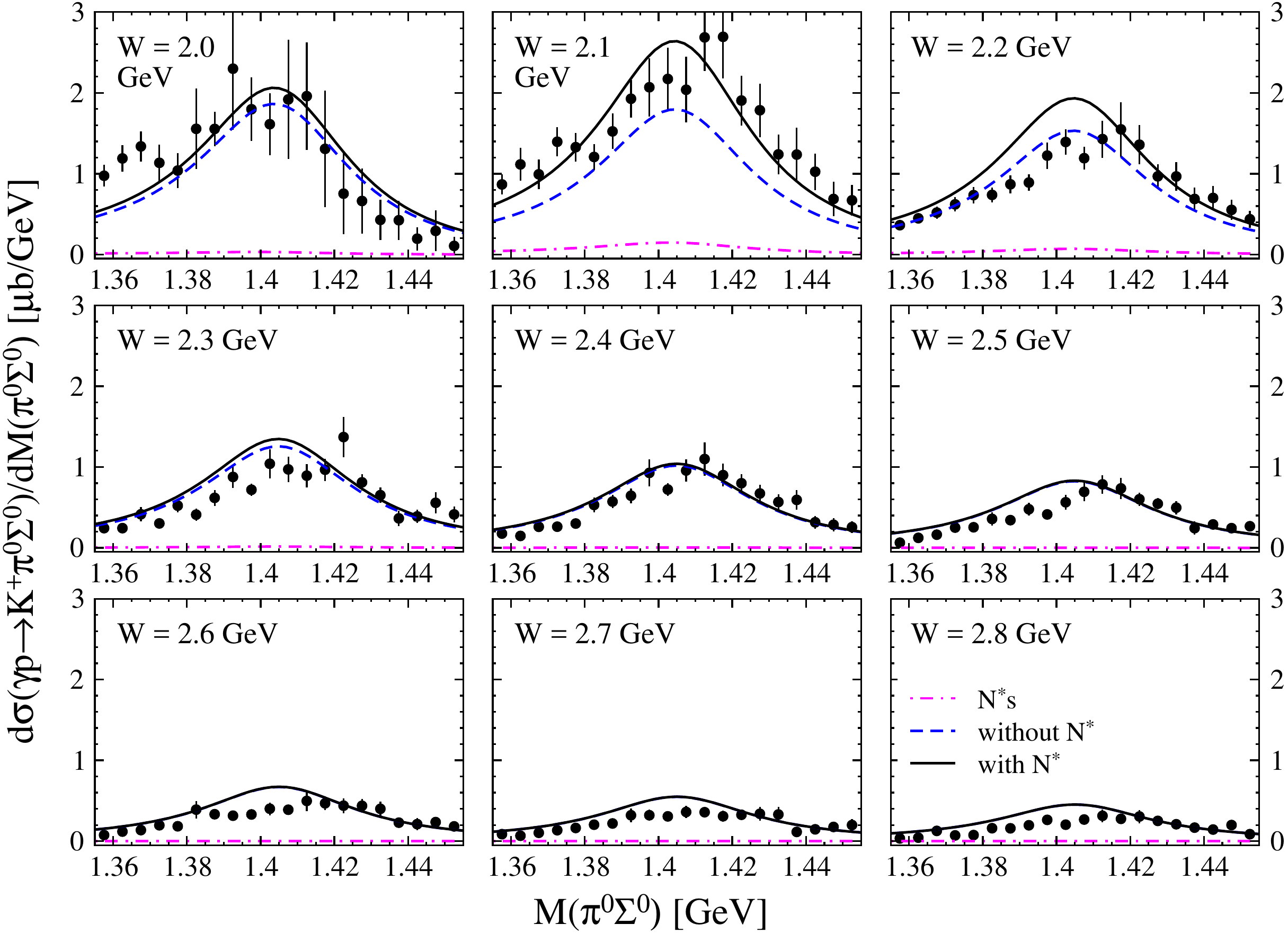}
\caption{(Color online) $d\sigma_{\gamma p\to K^+\pi^0\Sigma^0} / dM_{\pi^0\Sigma^0}$ for different cm energies $W=(2.0\sim2.8)$ GeV. The data are taken from Ref.~\cite{Moriya:2013eb}. The legends are the same with those for the left 
panel of Fig.~\ref{FIG3}.}
\label{FIG6}
\end{figure}

In Fig.~\ref{FIG6}, we draw the invariant-mass distributions $d\sigma_{\gamma p\to K^+\pi^0\Sigma^0}/dM_{\pi^0\Sigma^0}$ as a function of the  invariant mass $M_{\pi^0\Sigma^0}$ for different c.m. energies $W=(2.0 - 2.8)$ GeV, using Eq.~(\ref{eq:APP2}) and the numerical results for the two-body process, i.e., $\gamma p\to K^+\Lambda^*$, given in Fig.~\ref{FIG3}. The theoretical curves describe the CLAS data~\cite{Moriya:2013eb} qualitatively well for all the energies, as expected from Eq.~(\ref{eq:APP2}) by construction, due to the appropriate Breit-Wigner distribution and the strength from the two-body total cross section. From this observation, we find that the interference between the $\Lambda^*$ and other vector-meson productions, such as $K^*$, on the Dalitz plot is practically small as given in Ref.~\cite{Ryu:2016jmv}. In the low-energy region for $W\le2.2$ GeV, being similar to the total and differential cross sections as shown in Figs.~\ref{FIG3} and~\ref{FIG4}, the nucleon resonances make considerable contributions to the invariant-mass distribution as well, as expected. We also observe that the nucleon resonance contributions becomes almost negligible beyond $W=2.4$ GeV.

\begin{figure}[t]
\includegraphics[width=8.0cm]{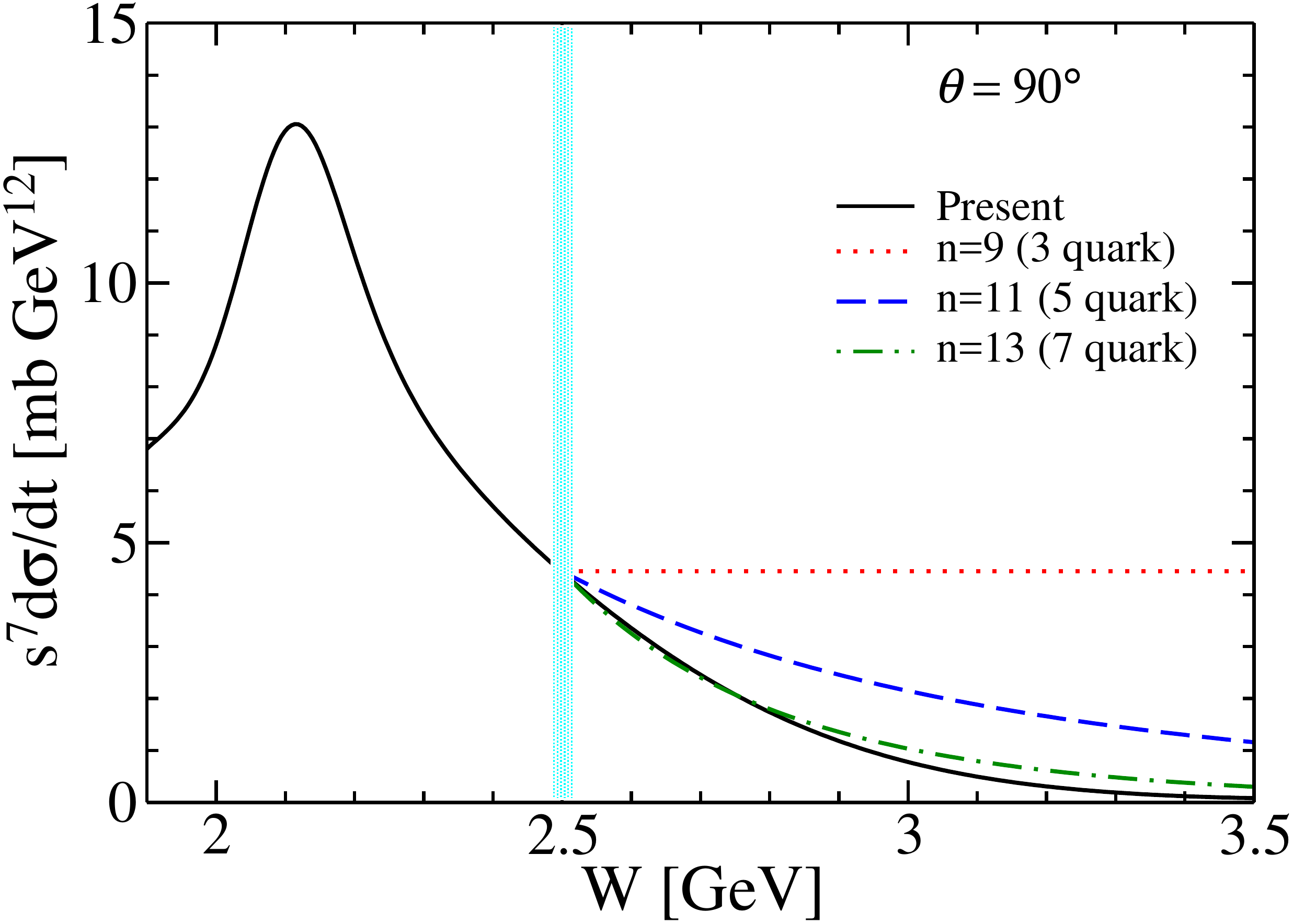}
\caption{(Color online) $s^7d\sigma_{\gamma p\to K^+\Lambda(1405)}/dt$ as a function of $\sqrt{s}\equiv W$ at the angle $\theta=90^\circ$ for the case for $n=(9,11,13)$.}       
\label{FIG7}
\end{figure}

Finally, we would like to test the CCR for the present reaction process. For this purpose, we take $\theta=\pi/2$ in the c.m. frame as done in Refs.~\cite{Chang:2015ioc,Kawamura:2013iia} and the numerical results are drawn in Fig.~\ref{FIG7}, where we draw $s^7d\sigma/dt$ as a function of $W$. The solid curve stands for the full result with the nucleon resonances. Following the CCR, if $\Lambda^*$ consists of three quarks as usual baryons, the resulting curve becomes flat (dotted), whereas it decreases with a tendency $\sim1/W^4$ for the five-quark $\Lambda^*$ (dashed) and $\sim1/W^8$ for the seven-quark $\Lambda^*$ (dot-dashed). Here, we assume that the resonance region is terminated at $W\approx2.5$ GeV as observed in the invariant-mass plot in Fig.~\ref{FIG6}. The curves obviously exhibit a decreasing behavior with respect to $W$, and the slope matches with $\sim1/W^8$, i.e., the seven-quark system for $\Lambda^*$. In Ref.~\cite{Chang:2015ioc}, the CLAS/Jlab experimental data was fitted with $[s^{n-2}/f(\theta)]\,d\sigma/dt$ as a function of $W$, where $f(\theta)$ denotes a scattering-angle dependent function, and it showed that $n=(6.5 - 11.3)$, depending on energies, indicating that $\Lambda^*$ is not a simple $uds$-quark state. Although the seven-quark state looks supported from the numerical results, it is still difficult to pin down the genuine structure of $\Lambda^*$ from this observation, because we need to take into account more realistic information on the structure functions, energy-angular dependences, and so on in the CCR as well as in our model. Hence, we can conclude that, more or less, $\Lambda^*$ is possibly distinctive from the simple $uds$-quark state, as far as we can reproduce the experimental data qualitatively well with the conventionally accepted model as in the present work.

\section{Summary}
\label{SecIV}
We have investigated the photoproduction of $\Lambda(1405)\equiv\Lambda^*$, i.e., $\gamma p\to K^+\Lambda^*$, by employing the effective Lagrangian with the Reggeized $t$-channel $K$ and $K^*$ exchanges at tree level. In addition to the ground state hadrons involved in the reactions process, we took five nucleon resonances near the reaction threshold into account. We performed numerical calculations for the energy and angular dependences of the cross section as well as the polarization observable, such as the photon beam asymmetry. The invariant-mass distribution for $\gamma p\to K^+\pi^0\Sigma^0$ was extracted from the two-body cross section with some reasonable assumptions. Finally, the internal structure of $\Lambda^*$ was explored by the constituent-counting rule (CCR), which provides the structure information of the hadrons in a reaction process in terms of the dimensional analyses of the scattering amplitudes at a large angle and high energies. Relevant results are summarized as follows:
\begin{itemize}
\item The nucleon resonances play an important role to reproduce the recent CLAS experimental data~\cite{Moriya:2013hwg} near the threshold region. Especially, the so-called PDG resonances, $N^*(2000,5/2^+)$ and $N^*(2100,1/2^+)$, provide considerable contributions to the total cross section $\sigma_{\gamma p\to K^+\Lambda^*}$, due to their relatively larger strong and EM couplings than other resonances, as far as we resort to available experimental and theoretical information to determine the couplings. 
\item As expected, angular dependence, such as $d\sigma_{\gamma p\to K^+\Lambda^*}/d\cos\theta$, is affected as well by the nucleon-resonance contributions in the low-energy region, whereas they get diminished as the energy increases. In overall, we observe strong forward-scattering enhancement, due to the $t$-channel $K$ exchange. The numerical results for the $d\sigma_{\gamma p\to K^+\Lambda^*}/dt$ are also given. We find that those results are affected by the inclusion of $N^*$ resonances only near the threshold region $W\le2.2$ GeV.
\item Assuming some reasonable conditions, i.e., no interferences between the hyperon and vector-meson resonances on the Dalitz plot for instance, we draw the invariant-mass plot for $d\sigma_{\gamma p\to K^+\pi^0\Sigma^0}/dM_{\pi^0\Sigma^0}$ in terms of the two-body cross section $\sigma_{\gamma p\to K^+\Lambda^*}$ and the partial decay width $\Gamma_{\Lambda^*\to\pi\Sigma}$. Qualitatively, the experimental data of the CLAS collaboration are reproduced well for the considered energy regions. From this observation, we find that the interference between the $\Lambda^*$ and $K^*$ productions on the Dalitz plot must be practically small and it is consistent to the result of Ref.~\cite{Ryu:2016jmv}. 
\item Finally, we examine the internal structure of $\Lambda^*$ via the CCR. The calculated curve of $s^7 d\sigma_{\gamma p\to K^+\Lambda^*}/dt$ as a function of the c.m. energy at $\theta=90^\circ$ shows a strong peaklike structure below $W\approx2.5$ GeV and a rapid decrease beyond it, due to the nucleon resonances and the Reggeized $t$-channel contributions, respectively. Using the most simple parametrization of the scattering amplitude via the CCR, the slope of the calculated curve beyond the resonance region matches roughly with that for $n=13$ from the relation $d\sigma/dt \propto 1/s^{n-2}$, which tells that $\Lambda^*$ is a seven-quark state.
\item However, because there are considerable theoretical uncertainties in this analysis, we can conclude safely that $\Lambda^*$ is possibly different from the usual $uds$-quark state as reported in Ref.~\cite{Chang:2015ioc}, if we take out results conservatively. Our numerical results for differential cross sections match the CLAS data pretty well even at a large angle ($\theta = 90^\circ$). Thus we believe that the results for $s^7 d\sigma/dt$ would be reliable to some extent even in relatively high-energy regions.
\end{itemize}

We want to mention that the photoproductions of the $\Lambda$, the $\Sigma$, and the  $\Sigma (1385)$ baryons have been already investigated extensively by using similar approaches in the literature, i.e., the so-called  Regge-plus-resonance framework~\cite{Corthals:2005ce,Corthals:2006nz,DeCruz:2012bv,He:2013ksa}. Meanwhile, the electroproduction of $\Lambda^*$ has its own interest, especially, as the CLAS collaboration recently has observed the  $\Lambda^*$ line shape in $\Lambda^*$ electroproduction~\cite{Lu:2013nza}. Although the relevant theoretical models appear in the literature~\cite{Williams:1992tp}, it is worth being studied within this effective model and related works are in progress. In summary,  the  present effective model approach with the Reggeized $t$-channel exchanges explains the $\Lambda^*$ photoproduction qualitatively well, showing good agreement with the CLAS data. As discussed above, however, it is still difficult to determine the genuine internal structure of $\Lambda^*$ within this simple model. Nonetheless, we have found a signal that $\Lambda^*$ can be distinctive from usual three-quark baryons. More realistic studies with sophisticated form factors and Regge treatments are in progress and will appear elsewhere. 

\section*{ACKNOWLEDGMENTS}
S.H.K. acknowledges support from the Young Scientist Training Program at the Asia Pacific Center for Theoretical Physics by the Korea Ministry of Education, Science, and Technology, Gyeongsangbuk-Do and Pohang City. The work of S.i.N. was supported by a Research Grant from Pukyong National University (2016). The work of D.J. was partly supported by a Grant-in-Aid for Scientific Research from JSPS (17K05449).
The work of H.-Ch.K. was supported by Basic Science Research Program through the 
National Research Foundation of Korea funded by the Ministry of Education, Science and  Technology (Grant No. NRF-2015R1A2A2A04007048). 
S.i.N.  also thanks J.~K.~Ahn, A.~Hosaka, H.~Kohri, and K.~H.~Woo for fruitful discussions.

\section*{APPENDIX: COUPLING CONSTANTS AND DECAY AMPLITUDES}
From the effective Lagrangians of Eq.~(\ref{eq:ResLag2}), the partial decay width of nucleon resonance $N^*(j^P)$ into $K\Lambda^*$ can be calculated as
\begin{eqnarray}
\Gamma\left[N^*(1/2^\pm) \to K\Lambda^*\right] =&
\frac{1}{4\pi} \frac{|\vec{q}|}{M_{N^*}} 
g_{K \Lambda^* N^*}^2 (E_{\Lambda^*} \pm M_{\Lambda^*}) ,                       
\cr
\Gamma\left[N^*(3/2^\pm) \to K\Lambda^*\right]=&
\frac{1}{12\pi} \frac{|\vec{q}|^3}{M_{N^*}} 
\frac{g_{K \Lambda^* N^*}^2}{M_K^2} (E_{\Lambda^*} \mp M_{\Lambda^*}) ,          
\cr
\Gamma\left[N^*(5/2^\pm) \to K\Lambda^*\right] =&
\frac{1}{30\pi} \frac{|\vec{q}|^5}{M_{N^*}} 
\frac{g_{K \Lambda^* N^*}^2}{M_K^4} (E_{\Lambda^*} \pm M_{\Lambda^*}).
\label{eq:Ampl1}
\end{eqnarray}
Here, the magnitude of the three-momentum and the energy for the $\Lambda^*$ in the rest frame of the nucleon resonance reads
\begin{eqnarray}
|\vec{q}|= \frac{1}{2M_{N^*}} 
\sqrt{[M_{N^*}^2 - (M_{\Lambda^*} + M_K)^2][M_{N^*}^2 - (M_{\Lambda^*} - M_K)^2]},
\,\,\,\,E_{\Lambda^*} = \sqrt{M_{\Lambda^*}^2 + |\vec{q}|^2}.
\label{eq:3-monen}
\end{eqnarray}

We need to know how the effective Lagrangians are related to the decay amplitudes to obtain the coupling constants. The decay amplitude for $N^* \to K \Lambda ^*$ can be expressed as follows~\cite{Oh:2007jd}: 
\begin{equation}
\label{eq:DA}
\langle K(\vec{q})\,\Lambda^*(-\vec{q},m_f) | -i \mathcal{H}_\mathrm{int} | 
N^*({\bf 0},m_j) \rangle                                               
=4 \pi M_{N^*} \sqrt\frac{2}{|\vec{q}|} \sum_{\ell,m_\ell} 
\langle \ell\, m_\ell\, \frac{1}{2}\, m_f | j \,m_j \rangle Y_{\ell,m_\ell} 
({\hat q}) G(\ell),
\end{equation}
where $\langle\ell\,m_\ell\,\frac{1}{2}\,m_f|j \,m_j \rangle$ and $Y_{\ell,m_\ell} ({\hat q})$ are the Clebsch-Gordan coefficient and spherical harmonics, respectively. The relation between the partial wave decay amplitude $G(\ell)$ and the decay width $\Gamma (N^* \to K \Lambda^*)$ can be derived as
\begin{eqnarray}
\label{eq:DA2}
\Gamma (N^* \to K \Lambda^*) = \sum_\ell |G(\ell)|^2 .
\end{eqnarray}
The spin and parity of the nucleon resonance place constraints on the relative orbital angular momentum $\ell$ of the final state.

Let us first consider the case of a $j^P = \frac{1}{2}^+$ resonance. The relative orbital angular momentum is constrained by angular momentum and parity conservation such that $\ell = 0$ is possible, i.e., only $s$ wave is allowed.  In a similar way, for the decays of  $j^P =(1/2^-,\,3/2^-)$ resonances  into $K \Lambda^*$, the final state is in the relative $p$ wave, and for  the resonances of $j^P = (3/2^+,\,5/2^+)$ in the relative $d$ wave. Finally, the $j^P =5/2^-$ resonance allowed only $f$ wave. Keeping this in mind, we can obtain the decay amplitudes in terms of the coupling constants for the decays of $j^P = (1/2^\pm,\,3/2^\pm,\,5/2^\pm)$ nucleon resonances into the $K \Lambda^*$ final state as follows:
\begin{eqnarray}
G\left(\frac{1-P}{2}\right) &=&
\sqrt{\frac{|\vec{q}|(E_{\Lambda^*} \pm M_{\Lambda^*})}{4\pi M_{N^*}}} 
g_{K \Lambda^* N^*}\,\,\,\,\mathrm{for}\,\,\,\,N^*(1/2^P),            
\cr  
G\left(\frac{3+P}{2}\right) &=& 
- \sqrt{\frac{|\vec{q}|^3(E_{\Lambda^*} \mp M_{\Lambda^*})}{12\pi M_{N^*}}}
\frac{g_{K \Lambda^* N^*}}{M_K}\,\,\,\,\mathrm{for}\,\,\,\,N^*(3/2^P), 
\cr
G\left(\frac{5-P}{2}\right)  &=& 
\sqrt{\frac{|\vec{q}|^5(E_{\Lambda^*} \pm M_{\Lambda^*})}{30\pi M_{N^*}}}
\frac{g_{K \Lambda^* N^*}}{M_K^2}\,\,\,\,\mathrm{for}\,\,\,\,N^*(5/2^P).
\label{R12}
\end{eqnarray}

\newpage
\end{document}